\documentclass[reprint,aps]{revtex4-1}
                   \usepackage{dcolumn}

\usepackage{amsmath}
\usepackage{graphicx}
\usepackage{bm}
\usepackage{color}

\newcommand{\GeV}{\makebox{ GeV}}
\newcommand{\beq}{\begin{equation}}
\newcommand{\enq}{\end{equation}}
\newcommand{\beqa}{\begin{eqnarray}}
\newcommand{\beqast}{\begin{eqnarray*}}
\newcommand{\enqa}{\end{eqnarray}}
\newcommand{\enqast}{\end{eqnarray*}}

\def\GeV{\nobreak\,\mbox{GeV}}

\providecommand{\U}[1]{\protect\rule{.1in}{.1in}}

\begin{document}

\title{  The interplay of hadronic amplitudes and Coulomb phase in LHC  measurements at 13 $\rm TeV$ }
\thanks{Corresponding author. Email: erasmo@if.ufrj.br\vspace{6pt}}  
\author{A. K. Kohara $^{\rm a}$ }
\author{E. Ferreira $^{\rm b} $ }
\author{M. Rangel $^{\rm b}$ }

\begin{abstract}
  Detailed analysis of the  measurements of   differential cross sections in the 
forward region of   pp elastic scattering at 13 TeV in LHC is performed. 
  The structures of the real and imaginary parts of the scattering amplitude  
are investigated, both requiring   exponential and linear factors.  Representations 
of the data in different conditions are compared, investigating the role of the 
Coulomb-nuclear interference and satisfying predictions from dispersion relations.   
The additivity of nuclear and Coulomb eikonal phases to determine the interference phase 
in pp scattering  is submitted to a comparison with assumption of phase equal to zero.  
The structures  of the real part of the scattering amplitude under the two assumptions 
are examined, and   direct comparison is  performed  with the information extracted   
from the data. The alternative criteria for the treatment of the interference phase 
lead to different values for the $\rho$ parameter between 0.1 and 0.13 while $\sigma$ 
differ by 0.18 mb,  and it is shown that the available data cannot discriminate the two 
choices. As an alternative study, we also present  the results for the amplitude 
  parameters and $ \chi^2 $ values when constant values of 2 and 3  radians are subtracted from 
the analytical expression for the interference phase.  
     \end{abstract}
    
\affiliation{ $ ^{\rm a}$ {\em Centre de Physique Th\'eorique, \'Ecole Polytechnique,
CNRS, Universit\'e Paris-Saclay, F-91128 Palaiseau, France } \\
$ ^{\rm b}$ {\em Instituto de F\'{\i}sica, Universidade Federal do Rio de Janeiro   \\
C.P. 68528, Rio de Janeiro 21945-970, RJ, Brazil }    }
 

\maketitle

\section { Introduction and formalism } 

  Measurements of elastic  pp scattering at 13 TeV \cite{totem_13} have been presented with 
  emphasis by the experimental group on  the determination of the $\rho$ parameter. 
We here analyse the  same  data with a fundamental framework, investigating the  
  structure of the real and imaginary parts of the complex elastic amplitude.
We  stress our view that in the analysis of elastic data 
 it is essential to obtain clearly  the identification of the real and 
imaginary parts of the complex amplitude 
\begin{equation} 
\label{amplitude}
  T(s,t)=T_R(s,t)+i ~T_I(s,t)~.
\end{equation}
 The quantitative, as much as possible model-free, description of the structure of the amplitudes 
provides  essential 
connection between  measurements and possible theoretical interpretation. The foundations of the 
strong and electromagnetic interactions (unitarity, causality, dispersion relations) deserve to be 
assumed as valid until clear deviation is imposed by    the data. In the laboratory,  measurements 
 are made of production rates $dN/dt$ and $d\sigma/dt$, while the identification of the two parts 
in the sum
 \begin{equation}
\label{cross}
  \frac{d\sigma}{dt}=\frac{d\sigma_R}{dt}+\frac{d\sigma_I}{dt}= (\hbar  c)^2 [(T_I)^2+(T_R)^2]  
\end{equation} 
is not at all trivial, depending on  analytical forms  and phenomenological models to be submitted to detailed 
 analysis.  This assertion seems obvious, but   it is important to avoid ambiguities and misunderstandings of 
essential points in the  interpretation of the data. 
   
The disentanglement of the two terms in the observed  modulus $d\sigma/dt$  is the 
crucial task. At each energy, parameterizations must   exhibit clearly  the 
properties of magnitudes, signs, slopes  and zeros of the real and imaginary parts.
Complementary support, as dispersion relations and connections with analyses at other 
energies, give  important   clues and control.  
  The intervention of the electromagnetic interactions must be treated coherently 
with a proposed  analytical form for the nuclear part, and the role of 
the phase of the Coulomb-Nuclear Interference (CNI) must be investigated.  
The   determination of scattering parameters require that analytical forms 
for the amplitudes be written explicitly and  checked for reliability.   

In the present study of forward data, each part of the amplitude is written with  an 
exponential factor with a slope, multiplying a linear   term in $t$, thus with 
three parameters. We consider that this is (almost) a model free construction, as these analytical 
forms are necessary and sufficient to describe the properties of the nuclear parts 
and yet are able to describe forward data with accuracy \cite{LHC_2017,LOW_ENERGIES}. 
The six parameters are studied  using  fits to data with appropriate statistical control. 
 The partial contributions to  the differential cross section are written in the forms 
\begin{eqnarray} 
\label{diffcross_eq_R}  
&&\frac{d\sigma_R}{dt}=\pi\left(\hbar c\right)^{2}  \times     \\
&&    \Big[\frac{\sigma(\rho-\mu_R t) }{4\pi\left(\hbar c\right)^{2}}~{{e}^{B_{R}t/2}
+F^{C}(t)\cos{(\alpha\phi)}\Big]^{2}}  \nonumber    
 \end{eqnarray}
and 
\begin{eqnarray}
\label{diffcross_eq_I}
&&\frac{d\sigma_I}{dt}=\pi\left(\hbar c\right)^{2}  \times      \\
&& \Big[\frac{\sigma (1-\mu_I t)}{4\pi\left(\hbar c\right)  ^{2}}~{{e}^{B_{I}t/2} 
+F^{C}(t)\sin{(\alpha\phi)}\Big]^{2}}  ~,   \nonumber
  \end{eqnarray}
where $t\equiv-|t|$,  $\alpha$ is the fine structure constant and 
$(\hbar c)^2=0.3894 $ mb $\GeV^2$. $F^C(t)$ and $\alpha\phi(t)$ account for 
 the proton form factor and phase of the Coulomb-nuclear  interference.  

 In Eqs.(\ref{diffcross_eq_R},\ref{diffcross_eq_I}) we have the  amplitudes 
\begin{equation}
\label{TR_TI_C}
 T_R(t)=T_R^N(t)+ T_R^C(t) ~ , ~~ 
  T_I(t)=T_I^N (t)+ T_I^C(t) ~  ,  
\end{equation}
 with  separate terms 
\begin{eqnarray}
 \label{TRN_TRC}
 &&  T_R^N(t)= \frac{\sigma (\rho -\mu_R t) }{4 \sqrt{\pi} \left(\hbar c\right)^2} ~  e^{B_R t/2} ~,  \\
 &&   T_R^C(t)=  \sqrt{\pi} F^C  \cos(\alpha \phi)  \nonumber
\end{eqnarray}
and 
   \begin{eqnarray}
\label{TIN_TIC} 
&& T_I^N (t)= \frac{\sigma ( 1 -\mu_I t )~}{4 \sqrt{\pi}\left(\hbar c\right)^2 } ~ e^{B_I t/2}  ~,  \\
&&  T_I^C(t)= \sqrt{\pi} F^C \sin(\alpha \phi)  ~.   \nonumber 
 \end{eqnarray}
 The  normalization (optical theorem) and $\rho$ parameter are   
\begin{equation}
 \sigma(s)= 4\sqrt{\pi}(\hbar c)^2 ~T_I^N(s,t=0)~~ , ~~ \rho = \frac{T_R^N(0)}{T_I^N(0)} ~. 
\label{sigma_rho} 
\end{equation} 
    With positive  $\rho$ and negative $\mu_R$ in pp at high energies \cite{LHC_2017}, there is a zero in the real 
amplitude, namely Martin's zero  \cite{Martin}, located at 
\begin{equation}
 t_R = \frac{\rho}{\mu_R}~,  
\label{tR_position}
\end{equation}
while, with negative $\mu_I$ the imaginary part points to a zero  with $t$ outside the forward range,
responsible for the dip in the differential cross section \cite{us_LHC}.  
 
The derivatives of the nuclear amplitudes at $t=0$   
\begin{eqnarray} 
 \label{BI_BR_eff}
 D_I=\frac{d}{dt}\log T_I^N (t)\bigg|_0= \frac{1}{2}[B_I-2 \mu_I]= \frac{1}{2} B_I^{eff} ~,  \\
 D_R=\frac{d}{dt} \log T_R^N (t)\bigg|_0= \frac{1}{2}[B_R-2 \frac{\mu_R}{\rho}]= \frac{1}{2} B_R^{eff} 
                             \nonumber
  \end{eqnarray} 
are related through the dispersion relations for slopes \cite{LOW_ENERGIES}. 
The average slope calculated  directly from $d\sigma/dt$  is the quantity 
\begin{equation}
\label{slope_dsigmadt}
  B= \frac{2}{(d\sigma/dt)\big|_0} [\frac{d}{dt}(d\sigma/dt)]\bigg|_0=  
  \frac{1}{1+\rho^2} \big [B_I^{eff}+\rho^2 B_R^{eff}\big]  ~. 
\end{equation} 
 The parameters   $\mu_I$ and $\mu_R$, with their roles of pointing  towards zeros in the amplitudes, 
are very important for the accurate description of the forward elastic data \cite{LHC_2017,LOW_ENERGIES}.
 
 We thus have the framework necessary   for the data analysis,
    with clear identification of the amplitudes and of the role of the six free  parameters 
   $\sigma$, $B_I$, $\mu_I$, $\rho$, $B_R$, $\mu_R$. 

  In the range of the data, the real part $d\sigma_R/dt$, that contains the $\rho$ parameter, is
about 1/100 of the imaginary part $d\sigma_I/dt$, and evidently the determination of parameters of the real 
part  requires  neat subtraction of the 99\%  magnitude  due to the imaginary part.  Although this separation 
is not explicitly identified in the fitting procedure exhibited in the experimental paper \cite{totem_13}, 
it occurs inside the fitting algorithm. Thus the value of $\rho$ informed  as "best fit"  results from 
a  delicate mathematical  separation, that must be clearly exhibited. There is a crucial role of the 
Coulomb interference phase in the extraction of  parameters, that we  investigate in detail. In particular, 
as we did in our previous analysis of LHC data at 7 and 8 TeV  \cite{LHC_2017}, to establish a reference, 
we  compare results  including the conventional interference  phase $\phi$,  based on the assumption of 
direct  addition of  the nuclear and Coulomb eikonal  phases,  with results obtained with phase put 
equal to zero. 

The assumption of additivity of eikonal phases in the superposition of interactions is 
believed to be successful in Glauber type calculations of hadronic collisions with nucleus, where 
addition  is made of interactions of similar nature (superposition of strong interactions of 
the incident hadron with the nucleons  of the nucleus). In the description of   pp elastic scattering
 the  interference occurs between  nuclear and electromagnetic forces that act on very different ranges. 
There is obvious possibility that in this case  the addition of eikonal phases is unrealistic, particularly 
at high energies. 
On the other hand, we can question about the relativistic derivation of the Coulomb phase in the 
pioneering works of Solov'ev \cite{Solovev} and  West-Yennie \cite{WY}, where the authors calculate the Coulomb 
phase using the simplest  Feynman diagrams. The results obtained provide phases with the same sign and 
similar magnitudes compared to the eikonal Coulomb phase. However it is important to stress that 
these derivations were in the pre-QCD era, without concept of quarks and gluons. Since the quarks 
couple with photons and gluons it is hard to separate strong and electromagnetic interactions, specially 
at high energies where the gluon density increases within the hadrons and nonlinear effects dominate 
the interactions. In purely hadronic  terms, Feynman diagrams of higher order include proton excitations 
($N^*$) and resonances ($\Delta$), and the pure additivity of separate electromagnetic and nuclear 
amplitudes is not satisfied.  
Thus there are complicated situations, and,  in a phenomenological treatment \cite{LHC_2017}, 
 in the present letter we  investigate the amplitudes entering in the calculation of $d\sigma/dt$  
with  interference phase  put equal to zero,  and also examine the influence of constant (non $t$-dependent) displacements in the values of the phase.

\section{ Results  } 

The results of the analysis of the data in the   $|t|$ range 0.0008 - 0.1996 $\GeV^2 $
are presented  in Table \ref{Table:Letter}. The headings of the table indicate the   
quantities determined in fits, namely the six parameters $\sigma$,  $B_I$, $\mu_I$,
$\rho$,$B_R$, and $\mu_R$. 
  A binned maximum likelihood fit is performed by MINUIT through the RooFit library 
available in the software toolkit ROOT 6.14/04 \cite{root}.  
  The analysis accounts  for statistical and systematic uncertainties and for correlations. 
 However, since the values of $\chi^2$ do not change much compared with the statistical 
uncertainties only, we show in the table  only  the statistical errors, that are needed for  the 
determination of the parameters of the  amplitudes. 
 
    As in the previous paper for 7 and 8 TeV \cite{LHC_2017}, we identify in the 13 TeV data the  
 real and  imaginary  amplitudes, and extract from the data the information on the parameters. The  
results for the real part are compared  with predictions 
for $\rho$  and for  amplitude derivative $D_R$ from dispersion relations \cite{LOW_ENERGIES}. 
Using the amplitudes written in Eqs.(\ref{TRN_TRC},\ref{TIN_TIC}), we compare calculations including  
Coulomb interference phase  and with phase put equal to zero. 
We also show the fitting parameters  assuming $\rho=0.131$ suggested  by dispersion 
relations based on usual parametrization of cross section data  \cite{LOW_ENERGIES} and putting 
zero phase. It is important to remark that in the inputs for the dispersion relations  we found 
no need for inclusion of odd terms, since the total cross section data are not sensitive 
to these contributions.
 We observe important differences in the  values obtained for the parameter $\rho$, while  
no significant differences between the $\chi^2$ values are found, but it is interesting that 
$\chi^2$ values tend to be smaller for zero phase. 
 
In order to isolate possible influences of mathematical nature, in the numerical work the Coulomb phase 
is treated with two different methods: using the implicit expression 
in Eq.(17) of the experimental paper \cite{totem_13}
that is called KL phase \cite{KL}, and using  appropriate  analytical expressions 
\cite{LHC_2017}, called KF phase ,  that are a generalization of Cahn's calculatiom \cite{Cahn} appropriate for 
amplitudes with independent exponential and linear factors. As shown in Table \ref{Table:Letter}, the values of parameters 
are practically the same
in the two cases, and this seems natural, since  the implicit interference phases are nearly the same 
for very small $|t|$, dominated by the term in $\log(-t)$ present in all phases based in the sum of eikonals. 
Thus numerically the  two methods of treating the calculation with interference phase (KL and KF) 
are here equivalent. 

There is an important difference of 0.16-0.18 mb  in the values of $\sigma$ obtained  with and without  
contribution of phase. This difference is only due to the pure  Coulomb contribution $T_I^C$  to  the 
imaginary amplitude $T_I$  in Eqs.(\ref{TR_TI_C},\ref{TIN_TIC}). 
The effects are obvious:  $ F^C \sin{\alpha \phi}$  is negative (while $T_I^N$ is positive), and its presence forces 
higher $\sigma$ to fit the data maintaining the same $d\sigma/dt$ at the origin. With higher value for $\sigma$
and  $d\sigma/dt$,
the value of  $\rho$  in $T_R$ must be smaller to fit the measured $d\sigma/dt$ near the origin. 
Thus: positive phase in $\sin{\phi}$ causes smaller $\rho$, and determination of $\rho$ goes influenced by 
the assumption of the presence  of the Coulomb-nuclear interference phase. 
 
We remark that  the value of the  derivative $D_R$  of the real part, given by $D_R$ in 
Eq.({\ref{BI_BR_eff}), is connected with $D_I$ and  satisfies  the prediction by the 
Dispersion Relations for Slopes \cite{LOW_ENERGIES}. 
  This indicates that  the true interference phase may be smaller than the calculation based on the additivity 
of eikonals.  Thus we observe  that the much discussed determination of the  value of the parameter 
$\rho$  as 0.1 is consequence of the assumption of the interference phase determined by 
the additivity  of nuclear and Coulomb eikonal phases. 
The question is not about   details in  calculation of the phase, but rather it is whether  the 
assumption of sum  of Coulomb and nuclear eikonal phases is justified in pp scattering at high energies. 
We show below that a test of this question  requires data  much more accurate than is  presently available. 
\begin{table*}[t]
\begin{center}
 \vspace{0.5cm}
 \small
\begin{tabular}{   c c c c c c c c c c}
\hline 
\hline 
 $\phi$&$\sigma$     &     $\rho$    & $B_{\rm I} $ &$ B_{\rm R} $ &$~\mu_{\rm R}$&$~\mu_{\rm I}$& $t_R$          &$B_I^{\rm eff}$&$\chi^2/$ndf\\
       &  (mb)       &               &$(\GeV^{-2})$ &$(\GeV^{-2})$ &$(\GeV^{-2})$ &$(\GeV^{-2})$ &$(\GeV^{2})$    & $(\GeV^{-2})$ &    =         \\
\hline
\hline
   \multicolumn{10}{c}{Condition I - all six  parameters free  }         \\ 
\hline
  KL &111.82$\pm$0.06&0.099$\pm$0.005&16.07$\pm$1.00&22.68$\pm$0.91&-3.78$\pm$0.37&-2.27$\pm$0.47&-0.026$\pm$0.004&20.61$\pm$1.20 &126.48/132=0.958  \\
\hline
  KF &111.84$\pm$0.06&0.097$\pm$0.005&16.13$\pm$1.33&22.72$\pm$1.19&-3.76$\pm$0.47&-2.26$\pm$0.62&-0.026$\pm$0.004&20.65$\pm$1.34 &126.83/132=0.961  \\
\hline
  0  &111.66$\pm$0.06&0.125$\pm$0.005&15.85$\pm$1.19&22.65$\pm$1.11&-3.84$\pm$0.44&-2.31$\pm$0.55&-0.033$\pm$0.004&20.47$\pm$1.22 &123.43/132=0.935 \\
 \hline  \hline 
 \multicolumn{10}{c}{  Condition II   - $\rho $ fixed by even signature dispersion relations  ; phase zero $  \phi=0 $   }         \\
      \hline
      \hline
0  &111.67$\pm$0.06&0.131 (fix)    &15.78$\pm$1.53&22.79$\pm$1.43&-3.96$\pm$0.53&-2.35$\pm$0.70&-0.033$\pm$0.004&20.48$\pm$1.60 &125.34/133=0.942  \\ 
 \hline\hline
\end{tabular}
 \caption{Results of fittings of the 138 points of the Totem measurements at 13 TeV 
in the $-t$ range from  0.0008 to 0.1996 $\GeV^2$. The Coulomb interference phase is 
  calculated according to Kundr\'at-Lokaji\~cek   \cite{KL}  and  KF \cite{LHC_2017}, and also phase is put equal 
to  zero. The quantity $B_I^{\rm eff}$  represents $B_I-2 \mu_I$ that is connected with the derivative $D_I$ of the imaginary 
amplitude at the origin according to Eq. (\ref{BI_BR_eff}) and is input for dispersion relation for  slopes. 
Attention must be given to the apparently small changes in    $\sigma$ (with and without phases), as they are   direct cause of the changes in $\rho$, according to explanation given in the text. 
\label{Table:Letter} }
\end{center}
\end{table*}

In  Fig.(\ref{TR_TC_details})  we show how  $\rho$ and phase affect the $|t|$  structure  of the 
real amplitude. We illustrate the  superposition of the quantities 
$T_R^N$  and $-T_R^C=-\sqrt{\pi} F^C \cos(\alpha \phi) $ that appear in the  real part $d\sigma_R/dt$ of 
the differential cross section. 
We remark that $d\sigma_R/dt$  is zero in the points where  $T_R^N$  and $-T_R^C $ cross each other, 
and this causes a  marked dip in $d\sigma_R/dt$. Depending on the value of $\rho$ the curves may not cross,  just 
approaching each other, with a marked reduction in  $d\sigma_R/dt$ in this region. The two situations are illustrated in the 
top and bottom parts plots of the figure.
In the RHS plots we show $d\sigma_R/dt$ for the two values of  $\rho$, without and with $\phi$, illustrated in the LHS.    
 \begin{figure*}[b]
  \includegraphics[width=8cm]{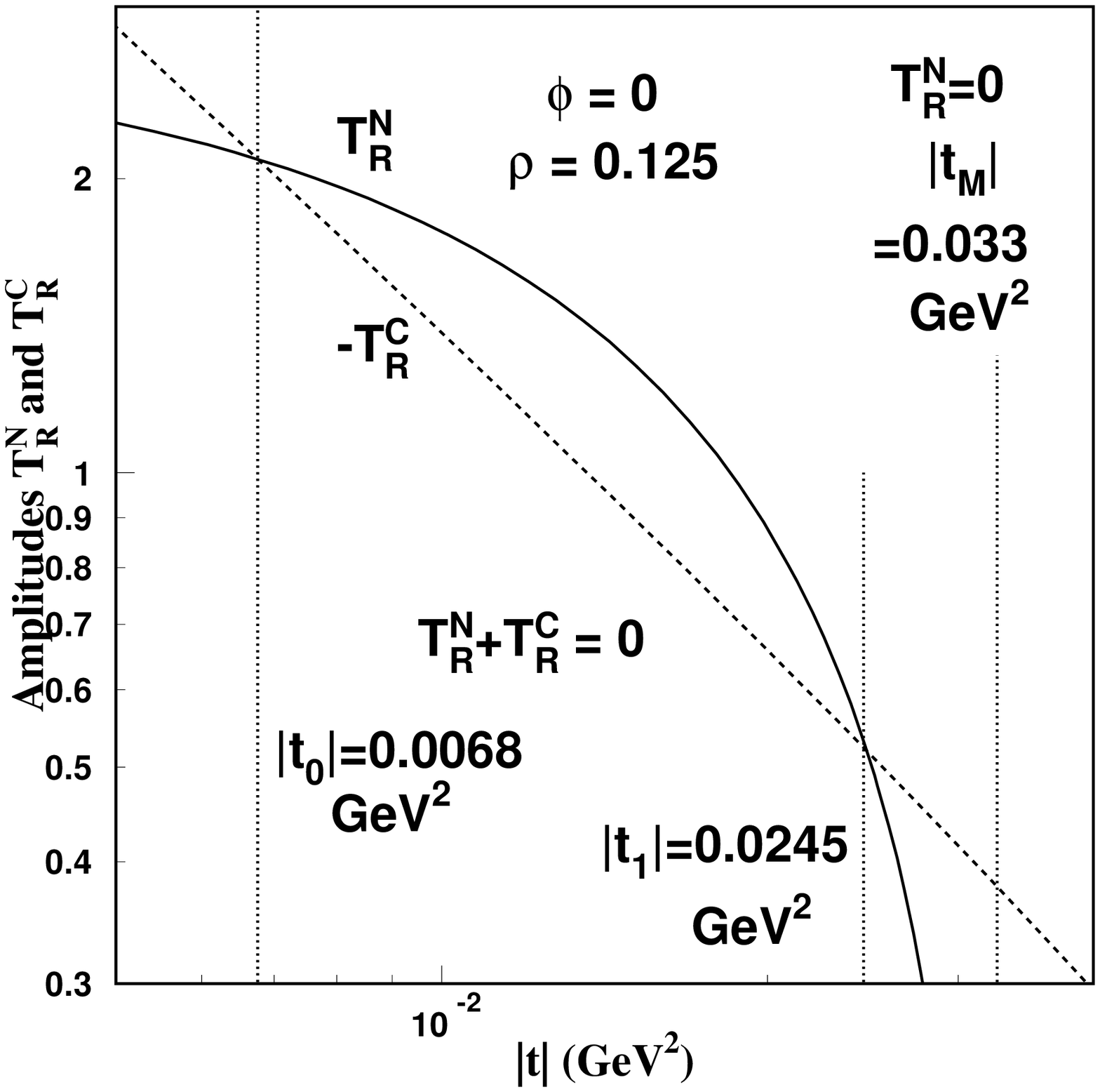}  
  \includegraphics[width=8cm]{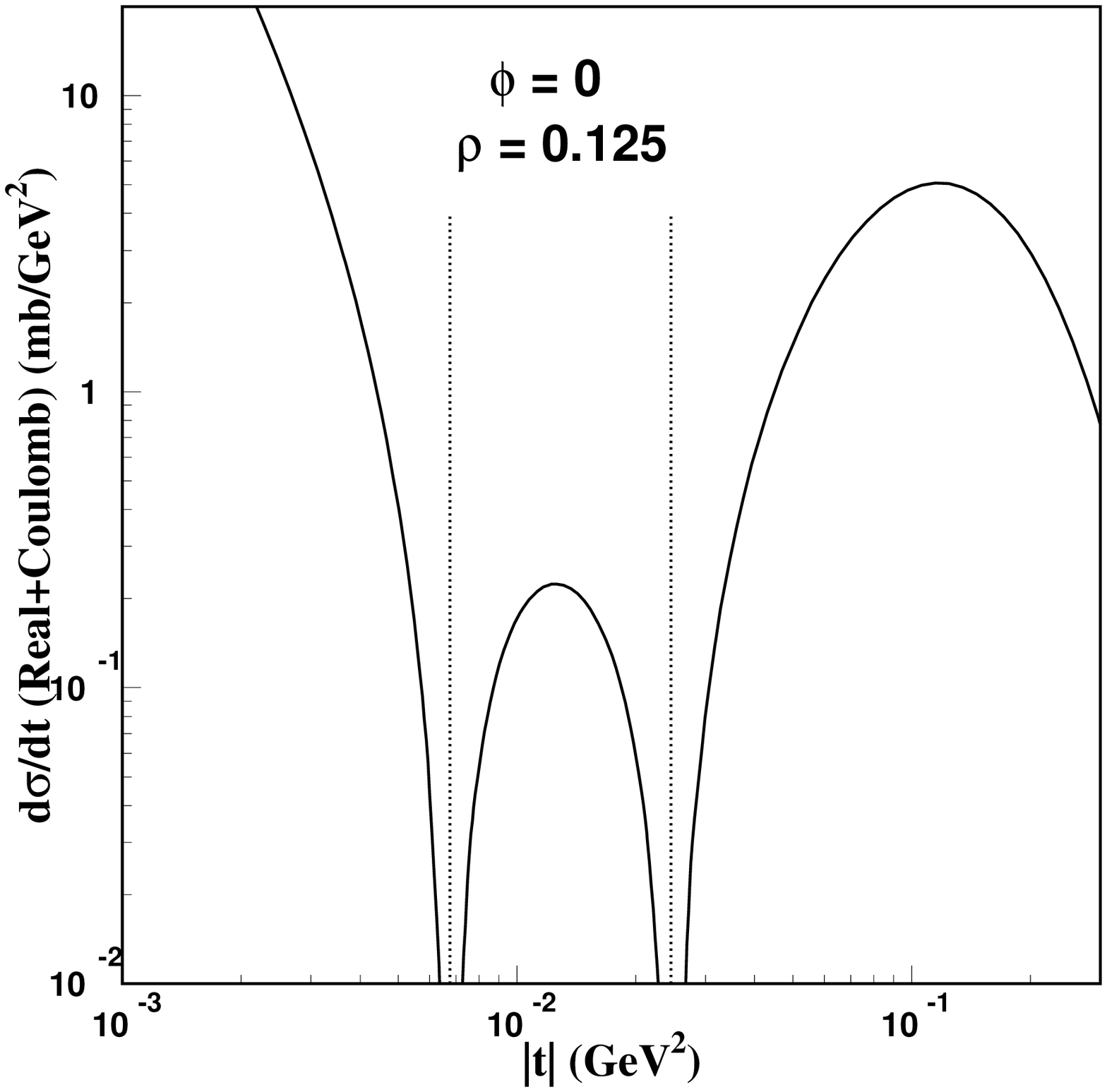}  
  \includegraphics[width=8cm]{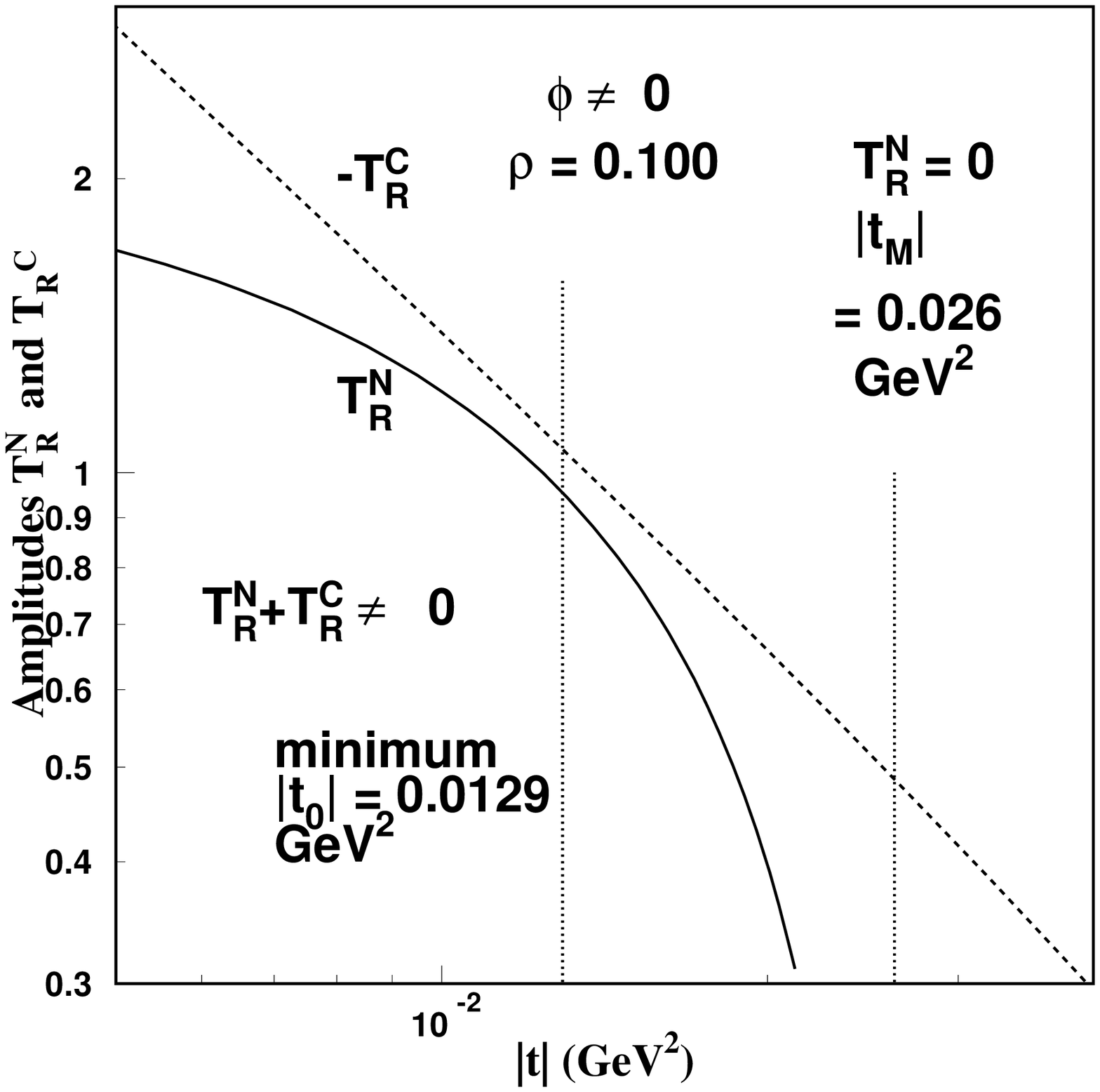}   
   \includegraphics[width=8cm]{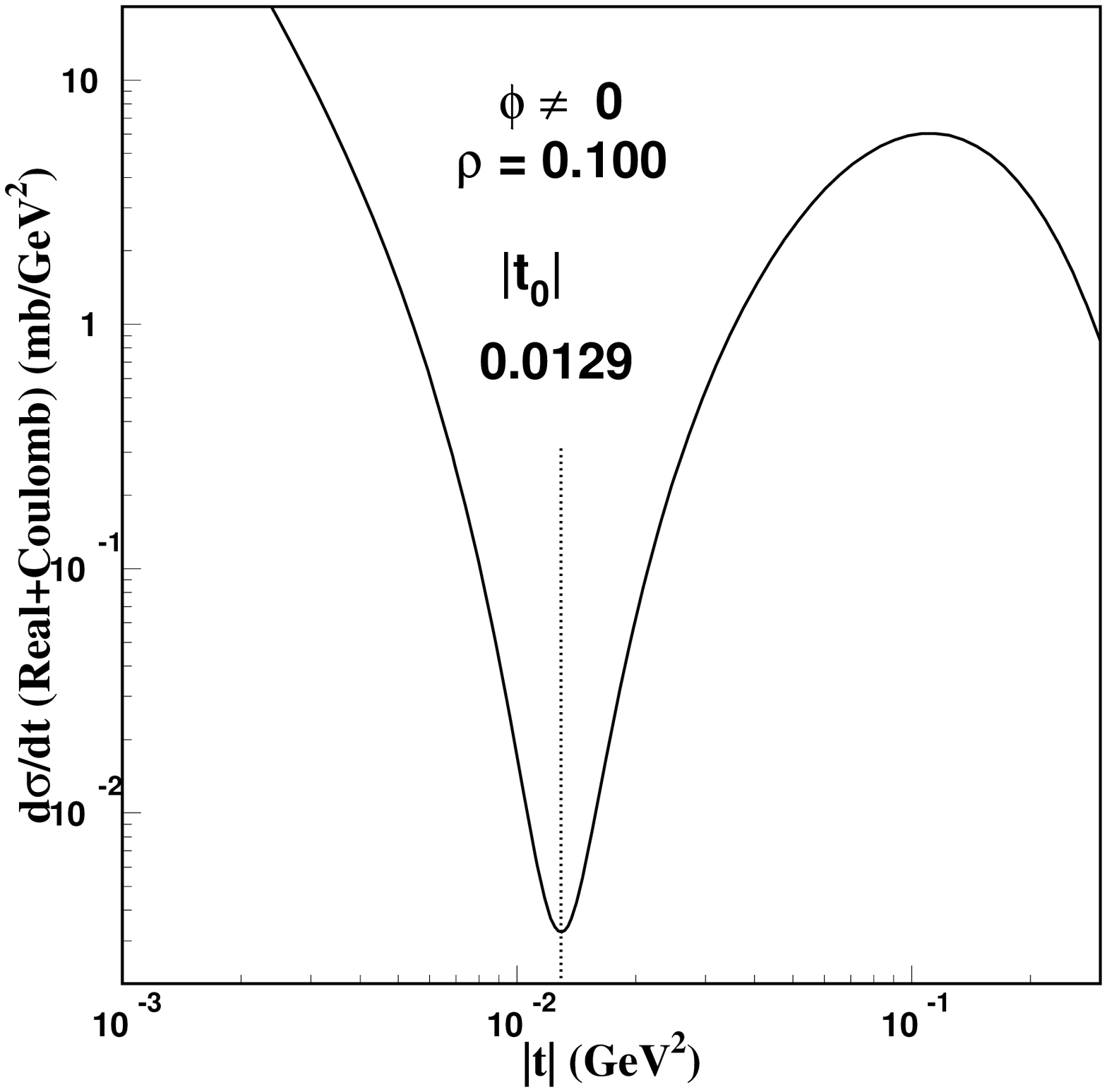} 
\caption{ Study of the properties of the real amplitude $T_R^N(t)$ and its superposition 
with the Coulomb amplitude $T_R^C(t)=\sqrt{\pi} F_C(t) \cos{\alpha \phi(t)}$
 in calculations  with and without  interference phase $\phi(t)$.   
Since $T_R^N$ is positive in the forward range, while $T_R^C$ is negative, the 
plots show the effects of the possible cancellations, or proximity, of the two quantities. 
The lines in the plots are calculated with the solutions given in Table \ref{Table:Letter} 
for the cases of phase zero  ($\rho=0.125$) and   phase non-zero KF ($\rho=0.100$).
Since the quantity $\alpha \phi$ is very small, the lines for  $T_R^C(t)$ are similar  
with and  without phase. On the contrary, the line for $T_R^N(t)$ in the plots  at the LHS, is
lower when the phase is present, because $\rho$ is smaller. The effect of the influence of 
the phase is dramatic. With zero phase, there are two zeros  in $T_R^N+T_R^C$, located at 
$|t_0|=0.0068 $ and $|t_1|=0.0245 ~ \GeV^2$; these cancellations  appear as dips in the 
quantity $d\sigma_R/dt$  as shown in the RHS at the top. 
On the other hand,  with the presence of the phase, with smaller $\rho$,  illustrated in the  
plots of the bottom, there is no cancellation in  the sum  $T_R^N+T_R^C$, but rather  an 
approximation of the $T_R^N$  and $-T_R^C$ lines, causing a minimum in  $d\sigma_R/dt$; this 
case is  illustrated in the RHS of the lower part od the figure. If the data are available 
with high precision, the two situations can  be distinguished. The reality with  the present data 
is shown in the next figure.} 
 \label{TR_TC_details}   
\end{figure*}

 To exhibit how the separation  of  real and imaginary parts appear in the data, 
we  introduce  in Fig.(\ref{RATIOS_13_TeV}) plots of  the ratio $T_R^2/T_I^2$ against $|t|$, where 
\begin{equation}
\frac{T_R^2}{T_I^2}=   \frac{~\big|T_R^N(t)+\sqrt{\pi} F^C(t) \cos(\alpha \phi)\big|^2}{~ \big|T_I^N (t)+\sqrt{\pi} F^C(t) \sin(\alpha \phi)\big|^2} ~. 
\label{Ratio_TR_TI}
\end{equation}
 In the figure we compare this quantity with the  points calculated with 
\begin{equation}
  \bigg[\frac{1}{\left(\hbar c\right)^{2}}\frac{d\sigma}{dt} - T_I^2 \bigg] \bigg/{T_I^2} ~, 
\label{Ratio_data}
\end{equation}
obtained from the $d\sigma/dt$ data with separation of the imaginary part.  
The ratio conveniently cancels most of the normalization uncertainty  of  $d\sigma/dt$. 
The figure shows  plots for the extracted quantity in the two investigated cases, and we 
observe  the enormous difficulty of the measurements to reach  a statistical precision   sufficient to 
distinguish $t$ dependences of the magnitude of the real part.  It is thus difficult to confirm the 
determination of  $\rho$  at 0.1, that  disagrees with the expectation from even signature  dispersion relations.  

\begin{figure*}[b]
  \includegraphics[width=8cm]{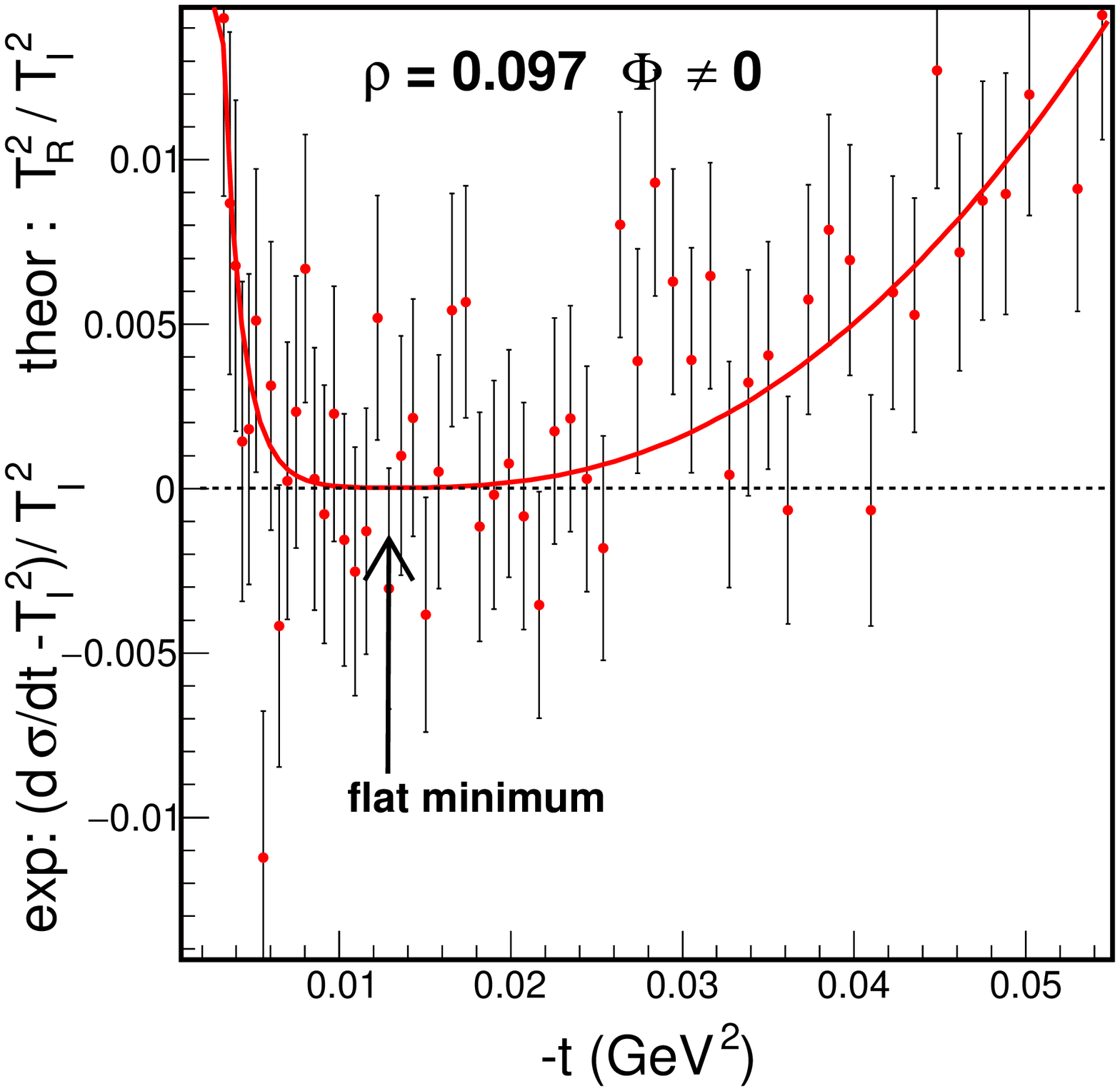}  
  \includegraphics[width=8cm]{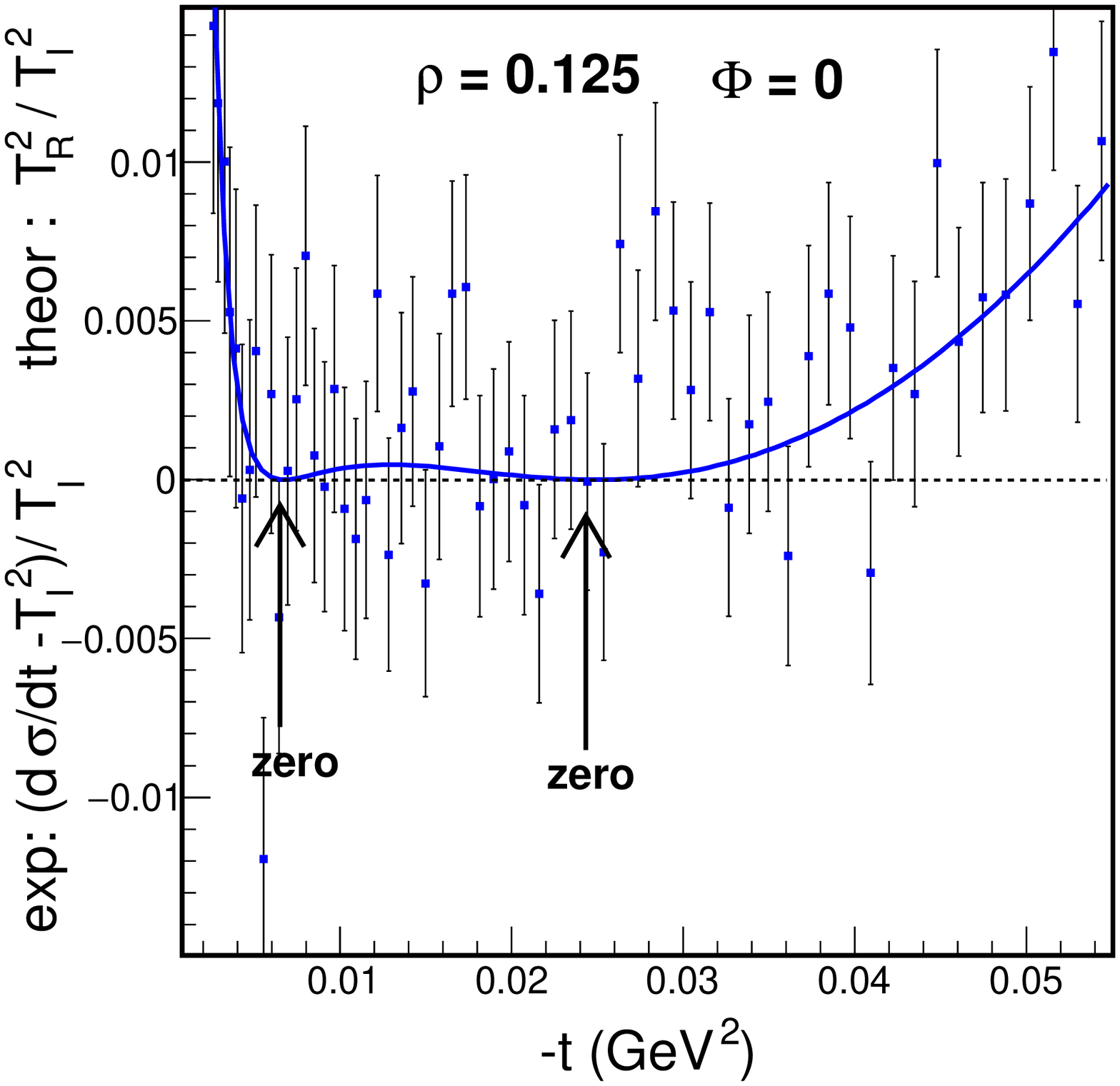}  
\caption{Study of the real part of differential cross sections of the 13 TeV data. 
The lines represent the squared ratios of the  parametrized forms of $T_R$ and $T_I$ 
given in Eq.(\ref{Ratio_TR_TI}).  The points are obtained as in Eq.(\ref{Ratio_data}): the 
data points of $d\sigma/dt$ minus the calculated imaginary background  $T_I^2$, divided  by the same.  
 } 
 \label{RATIOS_13_TeV} 
\end{figure*} 

 \subsection {Displacements by subtracting  constants in the phase  \label{displacements} }

\begin{table*}[t]
\begin{center}
 \vspace{0.5cm}
 \small
\begin{tabular}{   c c c c c c c c c c}
\hline 
\hline 
 $\phi$&$\sigma$     &     $\rho$    & $B_{\rm I} $ &$ B_{\rm R} $ &$~\mu_{\rm R}$&$~\mu_{\rm I}$& $t_R$          &$B_I^{\rm eff}$&$\chi^2/$ndf\\
       &  (mb)       &               &$(\GeV^{-2})$ &$(\GeV^{-2})$ &$(\GeV^{-2})$ &$(\GeV^{-2})$ &$(\GeV^{2})$    & $(\GeV^{-2})$ &    =         \\
\hline
\hline
   \multicolumn{10}{c}{Condition I - all six  parameters free  }         \\ 
\hline
 KF-2 &111.67$\pm$0.06&0.112$\pm$0.005&15.99$\pm$1.21&22.62$\pm$1.12&-3.74$\pm$0.45&-2.27$\pm$0.57&-0.030$\pm$0.004&20.53$\pm$1.54 &126.68/132=0.960  \\
\hline
 KF-3 &111.58$\pm$0.06&0.119$\pm$0.005&15.92$\pm$1.27&22.56$\pm$1.17&-3.73$\pm$0.47&-2.26$\pm$0.59&-0.032$\pm$0.004&20.47$\pm$1.63 &126.61/132=0.959  \\
  \hline  \hline 
 \end{tabular}
 \caption{ Results of fittings of the 138 points of the Totem measurements at 13 TeV 
in the $-t$ range from 0.0008 to 0.1996 $\GeV^2$. The Coulomb interference phase is 
  calculated according to   KF \cite{LHC_2017} subtracting  constant values  2 and 3 radians.
Attention must be given to the apparently small changes in 
$\sigma$, since  they are   direct cause of the changes in $\rho$.
\label{Table:Alternative} }
\end{center}
\end{table*}

\begin{figure*}[b]
  \includegraphics[width=8cm]{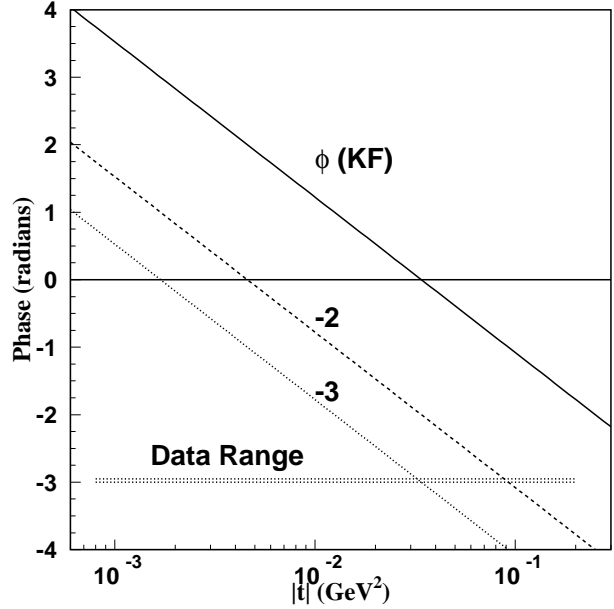}  
 \caption{ Plot of the  $t$ dependence of the phase $\phi$ (KF), together with representations of 
solutions with changes in the expression of the phase subtracting constant (no $t$ dependence) 
values of 2 and 3 radians. In the range of data, the phase  $\phi$ (KF) is dominated by a log(-X~t)  
term where $X$ depends on the parameters of the amplitudes and of the proton form factor. 
This term has a zero in the middle of the data range, with a maximum in $\cos{\phi}$. As explained 
in the text, the changes in $\rho$  are determined mainly by the changes in the imaginary 
part and in the total cross section, caused by differences  in $\sin{\phi}$.    
 \label{displacements-fig} }
\end{figure*}

\clearpage

\section{Remarks} 

With amplitudes given by Eqs.(\ref{TRN_TRC},\ref{TIN_TIC}), we analyse  the 13 TeV data  
 to investigate the interplay between Coulomb interference phase and the parameters of 
the amplitudes, with particular attention to  the    $\rho$ parameter of the real part. 
We stress  that $\rho$ and $\sigma$ are parameters fit to data using 
proper analytical forms of the amplitudes in the forward range, and   
are obtained   through  limits as $t \rightarrow 0$.

To investigate the role of the  Coulomb interference phase, 
we calculate with the usual construction  based on additivity of eikonal phases, 
  and   also obtain  results with phase put equal to zero.

In Fig.\ref{TR_TC_details} we exhibit the superposition of the hadronic and Coulombic  contributions 
to the real amplitude $T_R=T_R^N+T_R^C$  of Eqs.(\ref{TR_TI_C},\ref{TRN_TRC}) in the   low $|t|$ region, 
comparing the calculations with and without phase of Table \ref{Table:Letter}. Since $T_R^N$ is positive 
and $T_R^C$ 
is negative, they may cancel in two points (case of larger $\rho$=0.125) as  in the top part of the figure, 
or they just pass close (case of smaller $\rho$=0.100) as in the bottom part of the figure.  In the 
RHS of each case we show the effects of the two situations  on the real part of cross section  $d\sigma_R/dt$.  
In a very close scale there is a marked difference between the  two cases. 
 The results for the real part extracted from the data presented in Fig. \ref{RATIOS_13_TeV} show that 
the determination of  $\rho$ and the test of the phase influence  require  much  more data 
than  it is  available. To identify the dips or minimum in the data  for the real part it is necessary to 
improve the statistics and the regularity of  the original $d\sigma/dt$ data. 
  
To avoid doubts about the numerical procedures, as a test of the correct treatment of the influence of the phase, 
Table \ref{Table:Letter} shows   that as far as $\chi^2$ and parameter values are concerned, the fittings with 
 phase  KL 
treated using Eq.(17)  of the experimental paper \cite{totem_13}(where the phase is not represented by resolved 
 analytical explicit form but is evaluated in each step of the fitting procedure), and the phase  KF treated 
with the explicit calculation  \cite{LHC_2017}  are  equivalent.   
The table  shows   that, as far as $\chi^2$ values are concerned, the fits with zero phase 
may be  considered  equivalent  to those with  included phase, but the values of the  parameters are different.  
Even the calculation with fixed $\rho=0.131$ suggested by dispersion relations and zero phase is statistically 
equivalent to those with free $\rho$.  There is important difference in the first decimal digit in the values 
of $\sigma$, which is connected with the difference in the values of $\rho$. 

 It is remarkable that our previous analysis at 7 and 8 TeV  \cite{LHC_2017} presents similar features.

The imaginary part makes a very high contribution of 99 \%, so that  $d\sigma/dt$  must be measured with 
precision better than 0.1\% in order to discriminate between models with and without phase.
In particular, $\sigma$ must be obtained with error less than 0.1 mb, namely error in the second decimal
digit in mb.  
 
Values of $t$ where the interference with the Coulomb force has important local effects (the minimum and the dips
shown in Fig. \ref{TR_TC_details}), namely $|t|$  from 0.005 to 0.05 $\GeV^2$, appear  as  more important than 
a forward range  below 0.001 $\GeV^2$. This information may be useful for  the experimental effort.

To clarify with more detail the  connection  between the structure of the hadronic amplitudes and the phase of
the Coulomb-nuclear interference, in subsection \ref{displacements} we present
in Table \ref{Table:Alternative} and Fig.\ref{displacements-fig} results obtained with
changes in the expressions for the phase subtracting constant values 2 and 3 radians.

The parametrization of the amplitudes in equations (15) and (16) of the experimental paper \cite{totem_13}
   assumes  that real and imaginary parts 
run parallel in their   $t$ dependences. However, the ratio $T_R(t)/T_I(t)$ is not constant,  
dispersion relations for slopes  say that  $B_R \neq  B_I$,  $T_R$ has a zero
(Martin zero) interior to  the range of the   data.
The  parametrization of the  experimental  paper is not convenient also in the imaginary part, 
that behaves  unrealistically for higher $ |t|$, while it should point towards a zero around   0.5 $\GeV^2$.  

\bigskip

{\bf Importance of the real part at high energies and the theory for the interference phase } 

\bigskip

 The significance of the real part of the elastic amplitude  for possible signs of new physics 
    was pointed out   \cite{REAL_PART} in 2005, before LHC operation.
 The deviation  from conservative  dependence with the energy may signify violation of principles that 
 support dispersion relations, or induce new ideas in the theory of the strong interactions.
 The possibility of a small value $\rho=0.1$ at 13 TeV  brought renewed interest in the dynamics generated by
contribution of the odderon \cite{odderon} that was proposed a long time ago.
The Tel-Aviv group remarks  \cite{Tel-Aviv}  that contribution of odderon should be seen also 
at lower energies, and suggested that small $\rho$ can be investigated  through    
  screening (shadowing) and saturation effects in the gluon dynamics.

 The standard parametrization of the total cross section, without odd term,  in dispersion relations,
 provides  a simple and efficient framework for the description of elastic scattering, which should 
 not be abandoned before investigation of other possibilities for the phenomenology of elastic scattering.
Recalling the epistemological principle of Occam's razor (law of parsimony), it is not reasonable 
to look for a  sophisticated dynamical model without investigating simpler mechanisms. 
 
  The present paper  shows that the value of $\rho$ obtained from $d\sigma/dt$  may 
be consistent with the natural prediction 0.131 from dispersion relations, if the calculation  
of the Coulomb interference phase  is modified, assuming or not the additivity of the eikonal 
phases of nuclear and electromagnetic forces.
 
The success of Glauber calculation of the superposition of hadronic interactions in hadron-nucleus
collision is not necessarily a strong guideline   for the interference  of Coulomb  and strong interactions. 
The caution is particularly obvious at high energies, with participation of sea quarks, gluons 
and active intervention of  QCD vacuum in the nuclear part.  
  We do not know how  the electromagnetic field interferes with this 
nuclear dynamics.  We then point out that at high energies the theory for the phase based on the 
additivity in the eikonal representation of the nuclear and electromagnetic interactions, must be tested 
against experiment. 

The usual construction of the phase  since Solovev and West-Yennie  \cite{Solovev,WY,Cahn,Petrov2} is  
based on the Born approximation, with diagram of one-photon exchange. 
The  multi-photon exchanges   are seen to be important even at low energies
\cite{two_photon} and may modify strongly  the calculation of the interference of electric and nuclear 
forces.

On another hand,  the  proton form factor is now studied in terms of proton generalized structure 
functions \cite{ff} leading to the change in  the traditional parametrization of the dipole form factor. 
The influence  of the changes may apparently be not  very large in the very forward range, 
but we must observe that  in the study of the real part   any small influence may 
become important.

The important questions raised by the value of $\rho$ (violation of fundamental principles, 
screening effects in the gluonic dynamics, existence of odderon)  
should stimulate  more experiments of elastic pp scattering. Hopefully, richer data will be produced 
in  14 TeV  runs at LHC  that will   occur after two years from now.  
 
\begin{acknowledgments}
The authors wish to thank the Brazilian agencies CNPq and CAPES  
for financial support.  AKK wishes to thank  M\'arcio Taddei for useful comments concerning phases in quantum mechanics 
and L\'aszl\'o Jenkvoszky for the stimulating conversation during the workshop Hadron Physics 2018.  
\end{acknowledgments}

\end{document}